\documentstyle[aps,pra,epsfig,twocolumn]{revtex}
\begin{document}

\wideabs{

\title{Construction of exact solutions by spatial traslations in
  inhomogeneous Nonlinear Schr\"odinger equations.}

\author{Juan J. Garc\'{\i}a-Ripoll, V\'{\i}ctor M. P\'erez-Garc\'{\i}a and
  Vadym Vekslerchik\cite{email1}}

\address{Departamento de Matem\'aticas, E. T. S. I. Industriales,
  Universidad de Castilla-La Mancha, \\
  Avenida de Camilo Jos\'e Cela, 3, 13071 Ciudad Real, Spain.}

\date{\today}

\maketitle

\draft

\begin{abstract}
  In this paper we study a general nonlinear Schr\"odinger equation with a time
  dependent harmonic potential. Despite the lack of traslational invariance we
  find a symmetry trasformation which, up from any solution, produces
  infinitely many others which are centered on classical trajectories.  The
  results presented here imply that, not only the center of mass of the
  wave-packet satisfies the Ehrenfest theorem and is decoupled from the
  dynamics of the wave-packet, but also the shape of the solution is
  independent of the behaviour of the center of the wave.  Our findings have
  implications on the dynamics of Bose-Einstein condensates in magnetic traps.
\end{abstract}

\pacs{PACS number(s): 05.45.Yv, 03.75.Fi, 42.65.Tg}
}

\section{Introduction}

One of the most fruitful concepts of Physics is that of symmetries. From
high--energy physics to condensed matter, symmetries play a central role on our
understanding of the world.

As what concerns classical field theories, a symmetry is a transformation which
preserves the form of the equations. In this case the symmetry can help us in
many different ways. First of all, me may build solutions which have the same
symmetry as the equation. Let us take the two-dimensional spatially homogeneous
nonlinear Schr\"odinger (NLS) equation
\begin{equation}
\label{nlse}
i\partial_t\psi({\bf r},t) =
\left[-\hbox{$\frac{1}{2}$}\triangle + |\psi|^2\right]\psi({\bf r},t).
\end{equation}
This equation is invariant under spatial rotations, and therefore we are able
to search solutions with the given symmetry, $\psi =
\psi\left(\sqrt{x^2+y^2}\right)(x+iy)^n$.

Second and most important, Noether's theorem ensures us that once we have found
a certain symmetry in our model, it is possible to construct certain
quantities, often with physical relevance, which will be conserved during the
evolution. For instance, the invariance of Eq. (\ref{nlse}) under time
translations, spatial translations and rotations, give us seven conserved
quantities, which are the energy
\begin{equation}
E[\psi] = \int|\nabla\psi|^2 + \hbox{$\frac{1}{4}$}|\psi|^4,
\end{equation}
the linear moment of the center of mass
\begin{equation}
\label{nlse-p}
{\bf P}_c = \frac{d}{dt}\langle{\bf r}\rangle = \langle-i\nabla\rangle \equiv
\int -i\bar\psi\nabla\psi,
\end{equation}
and the angular momentum of the wavepacket
\begin{equation}
{\bf L} = \langle-i{\bf r}\times\nabla\rangle.
\end{equation}

Finally, the existence of symmetries and their associated conservation laws
helps us in the study of other properties. In particular, the invariance of Eq.
(\ref{nlse}) under Galilean transformations allows us to rewrite our equations
on an inertial frame of reference where the center of mass is still. This means
that the dynamics of the center of mass does not affect at all the dynamics of
other properties of the wavepacket.

In this paper we study a generalization of Eq. (\ref{nlse}) which lacks
traslational invariance. Nevertheless we will show a Galilean-like symmetry,
which allows us to construct, from any solution, a continuum of other ones
which follow different classical trajectories. We will show that this symmetry
implies a decoupling of the dynamics of the center of mass with respect to all
other properties of the wavepacket. We will also point out some very relevant
applications of our findings to the dynamics of Bose-Einstein condensates and
of several optical systems.

\section{The model}

In this work we will consider the following family of nonlinear Schr\"odinger
(NLS) equations with a general nonlinear term $G(|\psi|)$
\begin{equation}
\label{gpe}
i\partial_t \psi({\bf r},t) = \left[-\frac{1}{2}\triangle + V({\bf r},t) +
G(|\psi|)\right]\psi({\bf r},t).
\end{equation}
We will restrict our interest to the case of a quadratic potential $V({\bf
  r},t)$, i.e.
\begin{equation}
\label{cuad}
V({\bf r},t) = \frac{1}{2}\left({\bf r}, A(t){\bf r}\right),\quad
A_{ij} = \omega_i(t)\delta_{ij}.
\end{equation}

Eq. (\ref{gpe}) with potential (\ref{cuad}) is an accurate model of many
physical phenomena. In particular it describes the dynamics of a Bose-Einstein
condensate in the mean field approximation \cite{Dalfovo}, the propagation of
optical beams in graded index fibers \cite{momentos} and the propagation of
solitary waves in fiber trasmission lines with in-line phase modulators
\cite{Turitsyn}.

The nonlinear term, $G$, may adopt many different forms depending on the
particular application of Eq. (\ref{cuad}). The most classical cases are the so
called power nonlinearities $G(|\psi|^{2}) = \pm |\psi|^p$ which arise in
mean-field models with different spatial dimensionalities. In Nonlinear Optics
we also find many versions of the so called saturable nonlinearities, e.g.
$G(|\psi|^{2}) = \pm |\psi|^2/(1 + \beta |\psi|^2)$ as well as their Taylor
aproximations for small $u$, $G(|\psi|^{2}) = \pm |\psi|^2-\alpha |\psi|^4.$
But the nonlinearity need not be local, and in applications to Bose-Einstein
condensation one finds nonlocal expansions of the atom-atom interaction
$G(|\psi|^{2}) = \int K({{\bf r}}-{\bf{r}}') |\psi({\bf{r}}')|^2 d{\bf{r}}'$, where
the kernel is either radially symmetric $K(|{\bf r}-{\bf r}'|)$
\cite{Dalfovo,nolocal} or adopts more complex dependencies in the case of
dipole-dipole interactions \cite{Maciek}. These are only a few examples of the
many forms the nonlinear term may have.

The description of the dynamics involved in a NLS equation is of great interest
for applications. However, except for the very specific one dimensional case
with $G = \pm |\psi|^2, A = 0$, in which the equation may be integrated by
means of the Inverse Scattering method, nothing can be said about the structure
of the solutions. There are other tools such as the moment method which give us
information about the evolution of relevant integral quantities characterizing
the solution \cite{momentos}. In some cases, these methods are connected to the
conformal invariance of some classes of nonlinear Schr\"odinger equations
\cite{Gosh} but have several limitations: (i) They cannot be used to build
explicit solutions of the equations and (ii) they work exactly only on specific
cases. To derive a procedure which is valid for more general nonlinear problems
as the ones we consider here one must use some nontrivial approximations
\cite{SIAM-reson}.

In this paper we will be able to exploit the behavior of Eq. (\ref{gpe}) with
harmonic potential (\ref{cuad}) under spatial traslations to provide explicit
information on a whole class of time dependent problems as will be shown below.

\section{Building new solutions of the NLS by spatial traslations.}

\subsection{General case}
\label{prueba}

Let us consider a solution $\psi({\bf r},t)$ of Eqs. (\ref{gpe})-(\ref{cuad})
satisfying $\psi({\bf r},t= 0) = \xi({\bf r})$.  Our main result is that
given {\em any} solution, $\psi({\bf r},t)$, there exists a continuum of other
solutions which are of the form
\begin{equation}
\label{ansatz}
\psi_{\bf R}({\bf r},t) = \psi\left({\bf r}-{\bf R}(t),t\right)e^{i\theta({\bf r},t)},
\end{equation}
being ${\bf R}(t)$ and $\theta({\bf r},t)$ appropriate functions to be
determined later.

To check this point we proceed by inserting the ansatz $\psi_{\bf R}({\bf
  r},t)$ given by Eq.  (\ref{ansatz}) into Eq. (\ref{gpe}). Using the fact that
$\psi({\bf r},t)$ is a solution of Eq. (\ref{gpe}), we are able to cancel
several terms on both sides of the equation. If we impose that the new function
$\psi_{\bf R}$ be also a solution of Eq. (\ref{gpe}), we reach a solvability
condition which is made up of all the remaining terms
\begin{eqnarray}
 & & i\left(\nabla\theta - \frac{d{\bf R}}{dt},\nabla\psi\right) = \nonumber\\
 & & \hbox{$\frac{1}{2}$}\left[\partial_t\theta-i\triangle\theta + (\nabla\theta)^2
    + \left(2{\bf r}-{\bf R},A(t) {\bf R}\right)\right]\psi.
\end{eqnarray}

This is a set of partial differential equation for the unknown function
$\theta({\bf r},t)$.  Fortunately, it is possible to construct solutions by
choosing a linear phase,
\begin{equation}
\label{theta}
\theta({\bf r},t) = \left({\bf r},\frac{d{\bf R}}{dt}\right) + f(t),
\end{equation}
together with a trajectory ${\bf R}(t)$ determined by equations of Newton type
\begin{equation}
\label{newton}
\frac{d^2{\bf R}}{dt^2}  + A(t) {\bf R} = 0.
\end{equation}
By applying the hydrodynamic interpretation of the NLS equation
\cite{rotaciones}, the precise for of Eq. (\ref{theta}) leads to a
divergenceless velocity field, ${\bf v} = \nabla\theta = {\bf R}$, which is
responsible for the global displacement of the solution.

Finally, we need a global contribution to the phase, $f(t)$, which is
determined uniquely from
\begin{equation}
\label{eqf}
\frac{df}{dt} =
\left(\frac{d{\bf R}}{dt},\frac{d{\bf R}}{dt}\right) -
\left({\bf R},A(t) {\bf R}\right).
\end{equation}
This contribution can be calculated for each trajectory,
\begin{equation}
\label{fsol}
f(t) = \int_0^t \left[\left(\frac{d{\bf R}}{dt},\frac{d{\bf R}}{dt}\right) -
\left({\bf R},A(t){\bf R}\right)\right]dt.
\end{equation}

Therefore, what we get from Eqs. (\ref{ansatz}), (\ref{theta}), (\ref{newton})
and (\ref{fsol}) is a new solution of Eq. (\ref{gpe}) which is displaced from
the initial one.  It is remarkable that these explicit time-dependent solutions
are obtained by spatial traslations in a system which is not spatially
homogeneous and the dynamics is defined by simple, linear ordinary differential
equations.  This behavior is exclusive of the harmonic oscillator type
potential given by Eq. (\ref{cuad}) but it is not restricted to any specific
form of the nonlinear term or any dimensionality of the system.

\subsection{Evolution of stationary states}

A relevant type of solutions of Eq. (\ref{gpe}) are the so called solitary
waves or stationary solutions, which are of the form
\begin{equation}
\label{stat}
\psi({\bf r},t) = \phi_{\mu}({\bf r}) e^{i\mu t}.
\end{equation}
The existence and number of these solutions depends on the properties of the
nonlinear term. In this paper we will assume that the nonlinear term is such
that these solutions exist, which is in fact the case for most choices of $G$
of physical interest \cite{nolocal,Turitsynmat}. Then we may build from
(\ref{stat}) new solutions of the type
\begin{equation}
\phi_{({\bf R},\mu)}({\bf r},t) = \phi_{\mu}\left({\bf r}-{\bf R}(t)\right)
e^{i\left[\mu t + \theta({\bf r},t)\right]}.
\end{equation}
In this case the whole of the wavepacket moves following a classical orbit,
{\em while preserving the shape!} This interesting prediction can be confirmed
both experimentally and numerically.

\begin{figure}
  \begin{center}
    \resizebox{0.8\linewidth}{!}{\includegraphics{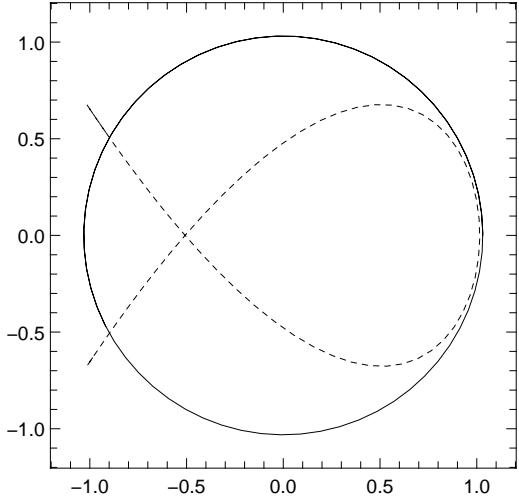}}
  \end{center}
  \caption{
    \label{fig-orbit}
    Trajectories of a solution which is initially stationary, and is suddenly
    displaced and imparted an initial velocity.  We plot the closed
    trajectories of the center of mass in the symmetric confinement (solid
    line, $\omega_x=\omega_y=1$, ${\bf R}(0)=(1,0)$, $\dot{\bf R}(0)=(0,1)$)
    and in the asymmetric trap (dashed line, $\omega_y=1.2$, $\omega_x=1$,${\bf
      R}(0)=(1,0)$, $\dot{\bf R}(0)=(0,1)$). The trajectories have been
    obtained integrating Eq. (\ref{gpe}) numerically.}
\end{figure}

In Fig.  \ref{fig-orbit} we show the evolution of two of such wavepackets,
first in the symmetric trap (solid line) and in the asymmetric trap (dashed
line). Such solutions were obtained by solving Eq. (\ref{gpe}) using a
split-step method on a Fourier basis with $128\times128$ modes. As our analysis
predicts, the shape of the wavefunction is preserved up to the numerical
precision of the computer.

\subsection{Addition of rotational terms.}

The proof presented in Sec. \ref{prueba} is also valid when the matrix $A(t)$
is non diagonal. An specific case of physical interest arises in Bose-Einstein
condensation when the trap which confines the atoms rotates. In that case it is
customary to study the system on the frame of reference which moves with the
trap, at angular speed $\Omega(t)$. On these coordinates the NLS equation reads
\begin{equation}
\label{gpe-rot}
i\partial_t \psi =
\left[-\frac{1}{2}\triangle + V({\bf r}) + G(|\psi|) + \Omega L_z\right]\psi,
\end{equation}
where $L_z$ is the Hermitian operator which represents the projection of the
angular momentum along the rotation axis and is given by
\begin{equation}
\label{newton-rot}
L_z\psi = -i({\bf r},J\nabla\psi).
\end{equation}
The antisymmetric matrix, $J$, is the generator of the rotations around the
$z$ axis
\begin{equation}
%J = \left(\begin{matrix}{cc}  0 & 1\\-1 & 0 \end{matrix}\right).
J = \pmatrix{0 & 1 & 0 \cr -1 & 0 & 0 \cr 0 & 0 & 0}.
\end{equation}

\begin{figure}
  \begin{center}
    \resizebox{0.9\linewidth}{!}{\includegraphics{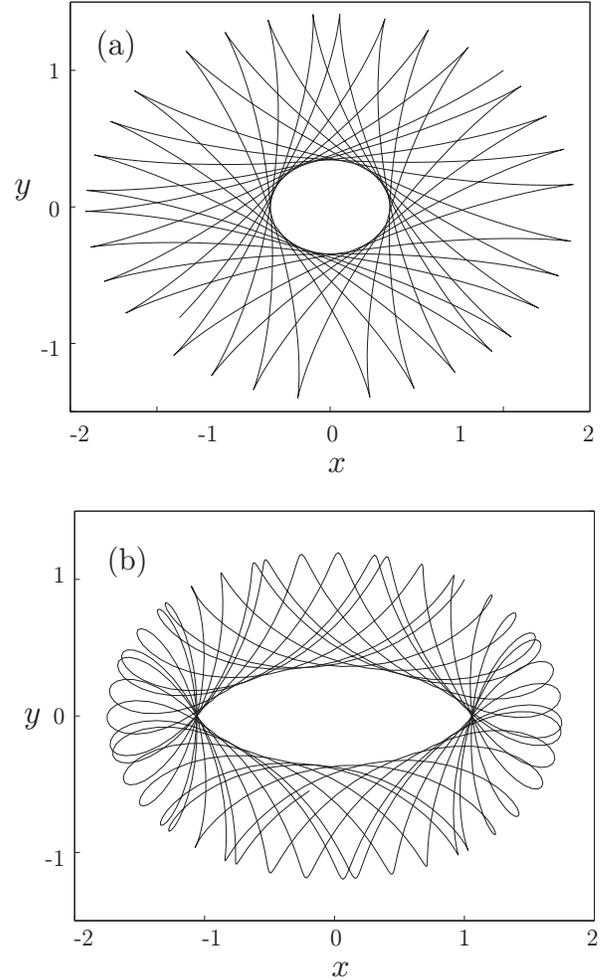}}
  \end{center}
  \caption{
    \label{fig-quasi}
    Trajectories of the center of a solution placed initially at $x(0) = 1,
    y(0) = 1,$ with $\dot{R}(0) = 0$ in two different situations: (a) $\omega_x
    = \omega_y = 1$, $\Omega = 1/2$, (b) $\omega_x= 1, \omega_y = 3/2, \Omega =
    3/2$.}
\end{figure}

By repeating the same calculations one arrives to a classical equation for the
wavepacket center, ${\bf R}(t)$, with an additional term due to the
centrifugal force
\begin{equation}
\label{withrota}
\frac{d^2{\bf R}}{dt^2} + \Omega J\frac{d{\bf R}}{dt} + A {\bf R} = 0.
\end{equation}
These equations form a linear system whose solutions are easy to obtain.
Specifically, for a two--dimensional oscillator and constant frequencies,
$\omega_x,\omega_y$ and $\Omega$, the solutions are quasiperiodic with
characteristic frequencies given by
\begin{eqnarray}
\label{omegas}
\omega_1^2  & = & \left|\sqrt{\left(\omega_x^2 + \omega_y^2\right)^2 +
\Omega^2\left(\Omega^2+2(\omega_x^2+\omega_y^2)\right)} \right.\nonumber\\
            & & \left. + \omega_y^2-\Omega^2-\omega_x^2\right| \\
\omega_2^2 & =& \left| \Omega^2+\omega_x^2 -\omega_y^2 \right.\nonumber \\
& + & \left. \sqrt{\left(\omega_x^2 + \omega_y^2\right)^2 +
    \Omega^2\left(\Omega^2+2(\omega_x^2+\omega_y^2)\right)} \right|.
\end{eqnarray}

Typical solutions are plotted in Fig. \ref{fig-quasi}, where the trajectory of
the wavepacket has been integrated numerically up from Eq. (\ref{withrota}).
It is important to stress the stability of these solutions: even in the case of
overcritical rotation ($\Omega > \omega_{x,y}$), when the centrifugal force
exceeds the restoring force of the harmonic potential, the motion is made of
bounded oscillations.

Incidentally, there is a formal equivalence between a NLS equation with a
rotating trap (\ref{gpe-rot}) with $\omega_x=\omega_y=\Omega$ and a
Ginzburg-Landau equation
\begin{equation}
  i\partial_t\psi =
  \hbox{$\frac{1}{2}$}(-i\nabla + {\bf A})^2\psi + |\psi|^2\psi,
\end{equation}
with uniform magnetic field, ${\bf A}^t = \Omega(y, -x, 0)$. In this
model it is particularly intuitive that the wavepacket should rotate around the
origin, due to the action of the uniform magnetic field, just as the above
discussed symmetry (\ref{newton-rot}) reveals.

\section{Decoupling the dynamics of the center of mass}

Up to now we have shown that given a solution $\psi({\bf{r}},t)$, we can build
many others, $\psi_{\bf R}({\bf{r}},t)$, by spatial traslations of the initial
data. The process can be reversed, so that given a wavepacket
$\phi\equiv\psi_{\bf R}({\bf{r}},t)$ which is a solution of Eq. (\ref{gpe}) we
can extract the dynamics of the center of mass, $\bf{R}(t)$, and the internal
dynamics of the wavepacket, $\psi(\bf{r},t)$.

The practical process is as follows. Let $\phi({\bf r},t)$ be any solution of
the NLS equation with a harmonic potential (\ref{gpe}). The center of mass
position is defined as
\begin{equation}
  {\bf R}_c(t) = \langle{\bf r}\rangle\equiv
  \int {\bf r} |\phi({\bf r},t)|^2 d^n{\bf r}.
\end{equation}
The dynamics of the center of mass, and of its associated momentum
(\ref{nlse-p}), is given by Ehrenfest's equations. Using the notation from
Quantum Mechanics, the expected value of an operator $A$ evolves according
to
\begin{equation}
  \label{heisenberg}
  \frac{d}{dt}A = \langle i[H(\psi),A]\rangle,
\end{equation}
where $H(\psi)$ is a nonlinear operator given by
\begin{equation}
  H = -\hbox{$\frac{1}{2}$}\triangle + V({\bf r},t) + G(|\psi|).
\end{equation}
Applying Eq. (\ref{heisenberg}) to ${\bf r}$ and to $(-i\nabla)$, we obtain the
following coupled ordinary differential equations
\begin{eqnarray}
  \frac{d}{dt}{\bf R}_c &=& \langle-i\nabla\rangle = {\bf P}_c,\\
  \frac{d}{dt}{\bf P}_c &=& \langle -\nabla V\rangle = -A{\bf R}_c.
\end{eqnarray}
With some manipulations it is easy to rewrite this system as a second order
differential equation
\begin{equation}
  \frac{d^2{\bf R}_c}{dt^2} + A{\bf R}_c = 0,
\end{equation}
with initial conditions
\begin{eqnarray}
  {\bf R}_c(0) &=& \int {\bf r} |\xi({\bf r})|^2 d^nr,\\
 \left. \frac{d{\bf R}_c}{dt}\right|_{t=0} &=& -i\int \bar\xi\nabla\xi d^nr,
\end{eqnarray}
where $\xi({\bf r})= \phi({\bf r},0)$ is the initial data of Eq. (\ref{gpe}).

This means that the center of mass already satisfies the equations for a valid
displacement in our symmetry transformation (\ref{ansatz}).  Hence we can
define a second wavefunction, $\psi({\bf r},t)$, which moves with the center of
mass, and which is the solution of Eq. (\ref{gpe}) with initial data
\begin{equation}
  \label{transformation}
  \psi({\bf r},0) = \xi({\bf r} + {\bf R}_c)
  \exp \left(-i{\bf r},\frac{d{\bf R}_c}{dt}\right).
\end{equation}
This second wavefunction, $\psi({\bf r},t)$, is located on the
center of mass
\begin{equation}
\int{\bf{r}}|\psi({\bf r},t)|^2 d^nr = 0,
\end{equation}
and it is the one that carries the dynamics of all observables ---widths,
angular momentum, circulation, etc--- completely free from the influence
of the center of mass.

Summing up, what all these transformations tell us is that {\em if we
  displace the initial data, or impart some speed to its center, we obtain the
  same solution}, $\psi({\bf{r}},t)$, centered on different trajectories.

This result has been obtained with the help of the Erhenfest theorem, which
states that the center of mass should satisfy an equation of Newton type, and
which was already known \cite{resonancias}. However, the result summarized in
Eq. (\ref{transformation}) is much stronger since it states that {\em the
  wavepacket is not affected by the dynamics of its centrum,} as this dynamics
can be integrated out of the equations.

\section{Application to the dynamics of the center of mass in Bose-Einstein condensates.}

Ever since the first works with dilute Bose-Einstein condensates, 
there has been an amazingly precise agreement between theory and experiments. From the
studies of normal modes, to the nucleation of vortices, it is usual to obtain a good
quantitative matching between the predictions (let it be collective frequencies or
critical speeds) and the actual measurements.

This is most intriguing in the case of experiments which involve a mechanical
perturbation of the condensate. We first focus on the study of the
collective excitations of a condensate. Such experiments consist of a periodic
modulation of the confinement of the condensate, and the subsequent study of
the oscillations of the wavepacket's widths. These manipulations have been
shown to not only modulate the widths, but to induce an exact, and extremely
strong resonance of the center of mass \cite{resonancias,torres}. Nevertheless,
both in the experimental results and in some rough models, the widths and the
center of mass seem to be decoupled, thus allowing us to precisely characterize
the normal modes of the condensate. That observed behavior is easy to
understand in the framework of the dynamics of displaced solutions described
here.

Another important application of Eq. (\ref{newton}) is the study of the center
of mass of the condensate in the regime of overcritical rotation, $\Omega >
\min\{\omega_x, \omega_y\}$.  In this regime the rotating condensate, which is
ruled by Eq. (\ref{gpe-rot}), suffers a centrifugal force which is stronger than
the restoring force due to the harmonic potential.  It is clear that in this
regime the condensate should be, and in fact it is found to be \cite{Madison},
untrapped.

However, the analysis of the eigenvalues of Eqs. (\ref{withrota}), which are
given by (\ref{omegas}) proves that the equilibrium point at $x=y=0$ is a
center and thus dynamically stable. Therefore, the only source of instability
for the condensate under overcritical rotations can be due to deformations of
the cloud.

This result is a bit more general than the one in \cite{zambelli}, where it is
proposed the existence of some configurations for the condensate, which
correspond to centered and elliptically deformed clouds that survive to the
action of the centrifugal motion. These configurations are stable under dipolar
perturbations (displacements of the cloud) and under quadrupolar excitations
(certain type of deformations). It remains an open problem to show whether such
states exist which are dynamically stable under {\em any} deformation.

As a side result which can be verified in experiments, a perturbed condensate
in a rotating trap suffers bounded oscillations around the origin with two
different frequencies, $\omega_1$ and $\omega_2$. These frequencies bear a
nontrivial dependency with respect to the angular speed of the trap
(\ref{omegas}), which can be used to better calibrate experiments. Finally we
must remark that the existence of two different oscillation frequencies for the
center of mass, $\omega_1$ and $\omega_2$, even in the symmetric trap
($\omega_x=\omega_y$) represents a splitting of the dipolar mode, which is
intuitively similar to the splitting of the quadrupolar mode due to the
presence of a vortex.

\section{Conclusions and discussion.}

In this paper we have built new solutions by simple time dependent traslations
in a system without traslational symmetry. It is remarkable, and probably a
special feature of the harmonic potential that this procedure works. Specially
striking is the case of traslation of stationary solutions whose center moves
harmonically whithout any distortion on the shape of the solution itself (only a
simple phase appears).

In relation with the previous finding, we have shown that the dynamics of the
center of mass is decoupled from the dynamics of all other properties of the
wavepacket.  This result stands on other works \cite{resonancias,torres}.
However, the contribution of this paper is different and stronger, since we
show that {\em motion of the center of mass can never influence any other
  properties of the wavepacket}, let it be a Bose-Einstein condensate or, in a
similar row, a solitary wave made of light. For all these systems the evolution
will be essentially the same, no matter the initial position and initial
velocity of the atomic cloud or solitary wave.

Our calculations are valid for any type of nonlinearity which is symmetric
under translations, and which depends only on the density, $|\psi|$. This
includes the cubic nonlinearity for Bose-Einstein condensates, $G =|\psi|^2$,
and most reasonable nonlocal terms \cite{nolocal}. This, and the fact that our
calculations do not depend on the dimensionality of the system, extends the
validity of this work to condensates with dipolar interactions, charged
condensates, light in Kerr media and light in saturable media.

The decoupling of the motion of the center of mass has also practical
consequences.  The invariance of the wavepacket dynamics up to displacements
and impulses on the initial data, explains why it is actually possible to
measure the frequencies of the normal modes of a condensate, even when the
center of mass of the condensate is known to be exponentially influenced by the
changes on the trapping potential \cite{resonancias,torres}.  This invariance
also benefits experiments with rotating condensates, as we have shown above,
and a simple analysis reveals an unexpected splitting of the dipolar mode of a
condensate.

The situation is different when other type of potentials are considered such as
stationary pinning potentials or any non-harmonic trapping potential, such as
some polynomial candidates, $V(x) \propto x^4$, which are being considered in
the context of all-optical condensation in very elongated traps.  The presence
of such potentials breaks our calculations, as the dynamics of the center of
mass couples to that of the widths by means of these external agents. We wonder
if this will imply some new and puzzling dynamics in future experiments.

\acknowledgements

This work has been supported by grant BFM2000-0521. V. Vekslerchik is
supported by Ministerio de Educaci\'on, Cultura y Deporte under grant SB99-AH777133.

\end{document}